\documentclass[journal]{IEEEtran} 				
\IEEEoverridecommandlockouts 						
\usepackage{amsmath,amssymb}
\usepackage{amsthm}
\usepackage{color}
\usepackage[mathscr]{eucal}
\usepackage{graphics,graphicx,multicol}
\usepackage{epsfig}
\usepackage{enumerate}
\usepackage{subfigure}
\usepackage{algorithmic}
\usepackage{algorithm}
\usepackage{morefloats}
\usepackage{enumerate}
\usepackage{subfigure}
\usepackage{multirow}
\usepackage{array}
\usepackage{pstricks, pst-node, pst-plot, pst-circ}
\usepackage{moredefs}
\usepackage{subfig}
\usepackage{courier}
\usepackage{tabularx}
\usepackage{multirow}

\begin{document}

\title{A Utility Proportional Fairness Radio Resource Block Allocation in Cellular Networks}
\author{Mo Ghorbanzadeh, Ahmed Abdelhadi, Charles Clancy\\
Hume Center for National Security and Technology\\
Virginia Tech, Arlington, VA, 22203, USA\\
\{mgh, aabdelhadi, tcc\}@vt.edu}
\maketitle

\begin{abstract}
This paper presents a radio resource block allocation optimization problem for cellular communications systems with users running delay-tolerant and real-time applications, generating elastic and inelastic traffic on the network and being modelled as logarithmic and sigmoidal utilities respectively. The optimization is cast under a utility proportional fairness framework aiming at maximizing the cellular systems utility whilst allocating users the resource blocks with an eye on application quality of service requirements and on the procedural temporal and computational efficiency. Ultimately, the sensitivity of the proposed modus operandi to the resource variations is investigated.
\end{abstract}


\providelength{\AxesLineWidth}       \setlength{\AxesLineWidth}{0.5pt}
\providelength{\plotwidth}           \setlength{\plotwidth}{8cm}
\providelength{\LineWidth}           \setlength{\LineWidth}{0.7pt}
\providelength{\MarkerSize}          \setlength{\MarkerSize}{3pt}
\newrgbcolor{GridColor}{0.8 0.8 0.8}
\newrgbcolor{GridColor2}{0.5 0.5 0.5}

\section{Introduction}\label{sec:intro}
The dramatic escalation of both mobile broadband subscriptions and their individual traffic, globally increased in Q4 2013 to $2$ - $31$ million additive subscriptions, $7800\%$ more data, and $250\%$ more voice traffic vis-\`{a}-vis Q1 2013 \cite{EricssonMobilityReport2013}, accompanied with the widespread prevalence of users' multi-service simultaneous deployment have been objects of grave concern for cellular providers. The situation is further compounded by the predictions estimating $3700$ million subscriptions by 2017 in retrospect to only $250$ million users in 2008 \cite{EricssonMobilityReport2013}. Moreover, smartphones' traffic diversity \cite{GhorbanzadehICNC2013}, arose from elastic (inelastic) traffic-generating delay-tolerant (real-time) applications, necessitates quality of service (QoS) requirements in order to elevate users' quality of experience (QoE), tightly bound to the subscriber churn \cite{QoS_3GPP}. As such, resource allocation researches attentive to the traffic diversity and 
dynamism have received significant interest.

Despite the fact that convoluted QoS-oriented rate assignment methods are devised mathematically, rarely are the germane modi operandi tailored to incumbent cellular infrastructures such as the Long Term Evolution (LTE), which pragmatically allocate resource blocks (RB)s \cite{QoS_3GPP} to user equipments (UE)s as opposed to precise (even optimal) fractional rates derived from complex bandwidth assigning schemes. In this paper, we study an RB assignment problem formulated as an integer nonlinear rate allocation optimization, whose positive integer solution represents RBs allocatable by LTE networks. In particular, we employ a Lagrangian relaxation \cite{Boyd2004} to solve an equivalent continuous dual optimization \cite{Wang2010} and utilize a boundary mapping of the continuous rates to extract RB candidates, those of which maximizing the cellular network's utility function can be allocated by the evolved Node B (eNB) \cite{QoS_3GPP}. Besides, we show that not only does this approach guarantee a minimum UE 
QoS due to its proportional fairness framework, but also it is able to allocate UEs the RBs temporally and computationally efficiently under a dynamic fashion. Unlike its simplicity, the algorithm is an effort to make RB-based communications' dynamic rate allocation pragmatically feasible.

In the spirit of providing relevant literature, subsection \ref{sec:related} surveys topical research papers about wireless networks' resource allocation, then it culminates by listing main contributions of the current article succinctly.

\subsection{Related Work}\label{sec:related}
Radio Resource allocation has been focused extensively in recent years. In \cite{Tychogiorgos2011,TychogiorgosGkelias2011}, the authors presented a non-convex optimization to maximize wireless networks utility functions heuristically to create a network stability. The authors of \cite{Harks2005} adopted a utility max-min fairness for the resource allocation in a network with elastic and inelastic traffic. In another work \cite{Tychogiorgos2012}, the authors suggested a utility proportional fairness for high signal-to-interference ratio wireless networks under a max-min architecture, contrasted their algorithm performance against certain traditional proportional fairness bandwidth allocation methods, and came up with a closed-form solution to seclude fluctuations. Methodologically similarly, \cite{Lee2005} cast a distributed power allocation in cellular systems with sigmoidal utilities and approximated a global optimal solution; however, their system utility maximization was at the expense of dropping users 
failing to warrant users a minimum QoS.

The authors of \cite{Ahmed_Utility3, Ahmed_Utility1, Ahmed_Utility2} developed a convex utility proportional resource assignment, using application utilities, which rendered priority to real-time applications. Neither did they consider temporal application usage, nor any user differentiation was included in the models. Analogously, \cite{GhorbanzadehMilcom2014} organized a utility proportional fairness to obtain cellular networks' optimal sector rates spectrally coexistent with other military communications. Finally, \cite{Tao2008} presented a subcarrier allocation in orthogonal frequency division multiplexed systems concentrating on delay constrained data and used network delay models \cite{GhorbanzadehICC2013} for the subcarrier assignment. The contributions of this paper are summarized below.

\begin{itemize}
\item We introduce an integer nonlinear optimization for wireless systems resource allocation where RBs are assigned to UEs with sigmoidal and logarithmic utilities modelling real-time and delay-tolerant applications respectively.
\item We demonstrate that the proposed RB allocation can be transformed into a convex rate assignment.
\item We explain that our algorithm prioritizes real-time applications above delay-tolerant ones and guarantees a minimum QoS via no user drops.
\item We solve the RB allocation effectively in terms of computation time and complexity by considering boundary-mapped discrete rates as potential RB candidates.
\end{itemize}

The paper remainder proceeds as follows. Section \ref{sec:Problem_formulation} presents the resource block allocation problem formulation. Section \ref{sec:sim} discusses a simulation setup, quantitative results, and discussions thereof. Section \ref{sec:conclude} concludes the paper.

\section{Problem Formulation}\label{sec:Problem_formulation}
\begin{figure}[!htb]
\begin{center}
\includegraphics[width=3.5in]{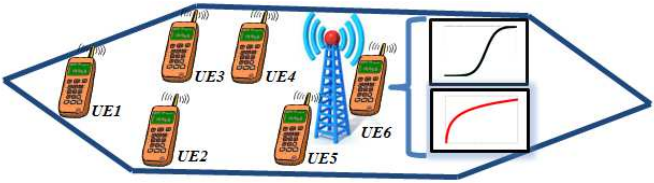}
\end{center}
\caption{System Model: eNB covers $M = 6$ UEs each running a delay-tolerant/real-time application modelled as a logarithmic/sigmoidal utility. The goal is for eNB to allocate RBs to UEs.} \label{fig:SysMod}
\end{figure}

Without loss of generality, consider a cell within a cellular infrastructure (Fig. \ref{fig:SysMod}) with an eNB covering $M$ UEs where the $i^{th}$ UE bandwidth and utility are respectively denoted as $r_{i}$ and $U_i(r_i)$. The latter indicates the $i^{th}$ user's satisfaction percentage for its rate $r_{i}$ and can be sigmoidal/logarithmic for a real-time/delay-tolerant application \cite{DL_PowerAllocation} as in equation (\ref{eqn:ApUtFn}), where $c_i = \frac{1+e^{a_{i}b_{i}}}{e^{a_{i}b_{i}}}$, $d_i = \frac{1}{1+e^{a_{i}b_{i}}}$, $r_{i}^{\text{max}}$ is a $100\%$ utility-achieving rate, and $k_i$ is the utility increase with enlarging $r_i$. \cite{DL_PowerAllocation} shows: 1) $U_{i}(0) = 0$ and $U_i(r_i)$ is an increasing function, implying utilities are non-negative and the higher assigned rate, the more QoS fulfillment. 2) $U_i(r_i)$ is twice differentiable and upper-bounded, implying utilities continuity. 3) $log(U_i)$ is concave.

\begin{equation}
\label{eqn:ApUtFn}
U_i(r_i) = \left\{
  \begin{array}{l l}
    c_i\Big(\frac{1}{1+e^{-a_i(r_i-b_i)}}-d_i\Big)\:\: ; \:\: \text{Sigmoidal}\\
    \frac{\log(1+k_ir_i)}{\log(1+k_i r_{max})}\:\: ; \:\:\text{Logarithmic}
  \end{array} \right.
\end{equation}

\subsection{Continuous Optimal Rates}
We cast a utility proportional fairness RB allocation optimization as equation (\ref{eqn:RB_Op}), where $R$ is the eNB bandwidth and $\textbf{r}$ = [$r_1,r_2,...,r_M$] vector is the rates assigned to the $M$ UEs. We aim at maximizing the objective function, referred to as the system utility, whilst maintaining a proportional fairness amongst individual application utilities. Interestingly, optimizations in the form of equation (\ref{eqn:RB_Op}), with no $r_i \in \mathbb{N}$ condition, are convex and have tractable global solutions \cite{Ahmed1}, not necessarily integers violating $r_i \in \mathbb{N}$. Adopting a Lagrangian relaxation \cite{Kiwiel2007} of $r_i \in \mathbb{N}$ in equation (\ref{eqn:RB_Op}) solves the so-called new problem easily by means of dual problem's Lagrange multipliers in equations (\ref{eqn:lagrangian}) and (\ref{eqn:dual_problem}), where $z_i \geq 0$ is the slack variable and $p$ is the Lagrange multiplier \emph{aka} price per unit bandwidth \cite{Kelly1998}.

\begin{equation}\label{eqn:RB_Op}
\begin{aligned}
& \underset{\textbf{r}}{\text{max}}
& & \prod_{i=1}^{M}U_i(r_i) \\
& \text{subject to}
& & \sum_{i=1}^{M}r_i \leq R\\
& & &  r_i \geq 0,\\
& & &  r_i \in \mathbb{N}, \;\;\;\;\; i = 1,2, ...,M.
\end{aligned}
\end{equation}

As mentioned before, the Lagrangian of equation (\ref{eqn:RB_Op}) is:

\begin{equation}\label{eqn:lagrangian}
\begin{aligned}
L(\textbf{r},p) = & \sum_{i=1}^{M}{\log(U_i(r_i))-p(\sum_{i=1}^{M}r_i + \sum_{i=1}^{M}z_i - R)}\\
                = &  \sum_{i=1}^{M}\Big({\log(U_i(r_i))-pr_i\Big)} + p(R-\sum_{i=1}^{M}z_i)\\
\end{aligned}
\end{equation}

And the dual problem of the Lagrangian is:

\begin{equation}\label{eqn:dual_problem}
\begin{aligned}
& \underset{p}{\text{min}} \: \underset{{\textbf{r}}}{\text{max}} \:L(\textbf{r},p)\\
& \text{subject to\quad} p \geq 0.\\
\end{aligned}
\end{equation}

The separability of $\sum_{i=1}^{M}({\log(U_i(r_i))-p r_i)}$ in $r_i$ legitimizes $\underset{\textbf{r}}\max \sum_{i=1}^{M}({\log(U_i(r_i)) - pr_i)} = \sum_{i=1}^{M}\underset{{r_i}}\max({\log(U_i(r_i))-pr_i)}$, hereby it transforms the aforementioned dual problem into the equation (\ref{eqn:dual_problem2}).

\begin{equation}\label{eqn:dual_problem2}
\begin{aligned}
& \underset{{p}}{\text{min}} \sum_{i=1}^{M}\underset{{r_i}}\max\Big({\log(U_i(r_i))-pr_i\Big)} + p(R-\sum_{i=1}^{M}z_i)\\
& \text{subject to\quad} p \geq 0.
\end{aligned}
\end{equation}

The Lagrange multiplier, i.e. $\frac{\partial \: \underset{{\textbf{r}}}{\text{max}} \:L}{\partial p}$, of the equation (\ref{eqn:dual_problem2}) yields in equation (\ref{eqn:dual_Shadow}), where $w_i = pr_i$ is the $i^{th}$ UE bid for resources and $\sum_{i=1}^{M}w_i = p\sum_{i=1}^{M}r_i$. Notably, $\sum_{i=1}^{M}z_i = 0$ minimizes equation (\ref{eqn:dual_Shadow}) where $\sum_{i=1}^{M}w_i = p_i R$.

\begin{equation}\label{eqn:dual_Shadow}
p = \frac{\sum_{i=1}^{M}w_i}{R-\sum_{i=1}^{M}z_i}
\end{equation}

In retrospect to what is being done thus far, we have divided the relaxed problem, equation (\ref{eqn:RB_Op}) excluding $r_i \in \mathbb{N}$, into two simpler optimizations in the UEs (equation (\ref{eqn:optUE})) and eNB (equation (\ref{eqn:optNB})) similarly to those in \cite{Ahmed_Utility1}. Calculated at the $i^{th}$ UE, the solution of equation (\ref{eqn:optUE}), i.e. $r_{i}(n) = \arg \underset{r_i}\max \Big(\log U_i(r_i) - p(n)r_i\Big)$, emerges from $\frac{\partial \log U_i(r_i)}{\partial r_i} = p(n)$ as the intersection of the horizontal line $y = p(n)$ and the curve $y = \frac{\partial \log U_i(r_i)}{\partial r_i}$ and is the optimal theoretical continuous rate for the $i^{th}$ UE.

\begin{equation}\label{eqn:optUE}
\begin{aligned}
& \underset{{r_i}}{\text{max}}
& & \log U_i(r_i) - pr_i \\
& \text{subject to}
& & p \geq 0\\
& & &  r_i \geq 0, \;\;\;\;\; i = 1,2, ...,M.
\end{aligned}
\end{equation}

\begin{equation}\label{eqn:optNB}
\begin{aligned}
& \underset{{p}}{\text{min}} \: \underset{{\textbf{r}}}{\text{max}} \:L(\textbf{r},p)\\
& \text{subject to\quad} p \geq 0.
\end{aligned}
\end{equation}

\subsection{Discrete Practical Rates (RBs)}
A utility proportional fairness is guaranteed in the solution $[r_{1}, ... , r_{M}] = arg \underset{{\textbf{r}}}{\text{max}}\prod_{i=1}^{M}U_i(r_i)$. $[r_1, ... , r_M]$ of the equations (\ref{eqn:optUE}) and  (\ref{eqn:optNB}). However, UE continuous rates are the global minimum of the relaxed optimization problem with $U({\textbf{r}}) = U(r_{1}, ... , r_{M}) = \prod_{i=1}^{M}U_i(r_i)$ system utility. For the sake of clarity and with no loss of generality, a two-UE cell with the system utility as a surface of rates $r_1$ and $r_2$ is displayed in Fig. \ref{fig:UtilityExample2D}. Despite the fact the a minimum of the convex surface (relaxed problem) is obtainable, here $U(54.2,64.5) =20$, a pragmatic resource allocation is meaningful provided that the allocative are RBs.

To obtain RBs, solving non-relaxed equation (\ref{eqn:RB_Op}), we initially leverage all possible discrete points in the $M$ dimensional domain of $U(r_1, ... , r_M)$ emerged from the cartesian product of possible values for the system utility. For example, $r_{ci}$'s gamut ranging from $1$ (proportional fairness requires a nonzero allocation) to $Q$, typically of the order of a hundred, generates $Q^{M}$ RB possibilities. Nonetheless, such as exhaustive RB search grows computationally complex for large systems; as such, we focus on neighbor discrete points of the continuous optimal rates, i.e. lower (upper) integer points less (more) than the continuous rate. Despite its simplicity, this approach decreases the $Q^{M}$ RB possibilities to $2^{M}$ ones, enduring less computational complexity. In essence, this corresponds to computing $[\lfloor r_{1} \rfloor, ... , \lfloor r_{M} \rfloor]$ and $[\lceil r_{1} \rceil, ... , \lceil r_{M} \rceil]$ and account for combinations thereof. For example, while there are 
many valid discrete points on the abscisses in $U(r_1,r_2)$ domain in Fig. \ref{fig:UtilityExample2D}, leveraging $[r_1,r_2]$ boundary points engenders $\{(\lfloor r_{1} , \lfloor r_{2}), (\lceil r_{1} , \lfloor r_{2}), (\lfloor r_{1} , \rceil r_{2}), (\lceil r_{1} \rceil r_{2})\}$ (red, green, yellow, and cyan points in Fig. \ref{fig:UtilityExample2D}), the closest integers to the continuous rates $r_1$ and $r_2$ (blue). Consequently, the search space constricts to four points alone vs. all possible discrete points on the surface of the system utility.

Despite the proximity of the preceding RBs to the optimal continuous rate vector, they may not fulfill the original relaxed problem's constraint $\sum_{i=1}^{M}r_i \leq R$ due to the rounding effect. Henceforth, we test candidate RBs and opt out those fulfilling the mentioned requirement. Such a confinement is influential stiff in high dimensional systems as it eliminates violating RBs juxtaposed to the continuous rates. Ultimately, we evaluate the system utility at the sifted RBs and store the utility maximizing ones as feasible candidates. The RB allocation procedure (equations (\ref{eqn:optUE}) and (\ref{eqn:optNB}) and boundary point mapping) is illustrated in Algorithm (\ref{alg:Our_UE}) and (\ref{alg:Our_eNodeB}).

\begin{algorithm}
\caption{UE Algorithm (from \cite{Ahmed_Utility2})}\label{alg:Our_UE}
\begin{algorithmic}
\STATE {Send initial bid $w_i(1)$ to eNB}
\LOOP
	\STATE {Receive shadow price $p(n)$ from eNeB}
	\IF {STOP from eNB} %
	\STATE {Calculate allocated rate $r_i ^{\text{opt}}=\frac{w_i(n)}{p(n)}$}
			\ELSE
	\STATE {Calculate new bid $w_i (n)= p(n) r_{i}(n)$}
	\IF {$|w_i(n) -w_i(n-1)| >\Delta w(n)$} %
	   	\STATE {$w_i(n) =w_i(n-1) + \text{sign}(w_i(n) -w_i(n-1))\frac{l_3}{n}$}
	   	\COMMENT {$\Delta w = l_1 e^{-\frac{n}{l_2}}$ or $\Delta w = \frac{l_3}{n}$}
	\ENDIF
	\STATE {Send new bid $w_i (n)$ to eNB}
		\ENDIF
\ENDLOOP
\end{algorithmic}
\end{algorithm}

\begin{algorithm}
\caption{eNB Algorithm}\label{alg:Our_eNodeB}
\begin{algorithmic}
\LOOP
	\STATE {Receive bids $w_i(n)$ from UEs}
	\COMMENT{Let $w_i(0) = 0\:\:\forall i$}
			\IF {$|w_i(n) -w_i(n-1)|<\delta  \:\:\forall i$} %
	   		\STATE {STOP and calculate rates  $r_{i}^{\text{opt}}=\frac{w_i(n)}{p(n)}$ }
		\ELSE
	\STATE {Calculate $p(n) = \frac{\sum_{i=1}^{M}w_i(n)}{R}$}
	\STATE {Send new shadow price $p(n)$ to all UEs}
	\ENDIF
\ENDLOOP
\COMMENT{Steady sate continuous rate is found.}
\LOOP
	\STATE {Continuous value of $r_i$ is $r_i^{opt}$.}
    \COMMENT {Map the floors of continuous rates less than unity to one unit.}
    \STATE {Calculate the ceilings and floors of the $r_i$.}
\ENDLOOP
    \STATE{List all possible sequences that can be obtained from the floors and ceilings.}
    \IF {sum of the discrete rates surpasses $R$}
        \STATE {Eliminate that sequence of discrete rates.}
    \ELSE
        \STATE {Store that sequence of discrete rates as a candidate for RB allocation.}
    \ENDIF
\LOOP
	\STATE {Calculate the system utility for the stored RB candidates.}
    \STATE {Store RBs which maximize the utility.}
\ENDLOOP
\end{algorithmic}
\end{algorithm}

It is remarkable that Algorithm (\ref{alg:Our_eNodeB}) always maps less-than-unity continuous rates to unities as the minimum possible assignment under a proportional fairness policy must be nonzero \cite{Kelly1998}. Many services not requiring many resources will choose one RB. Furthermore, our allocation paradigm may result in numerous RB sequences per continuous rate vector rendering eNBs a flexibility to distributing the resources. This can be of significant consequence cellular network with several eNBs because it can reduce inter-cell and intra-cell interferences by letting eNBs allot UEs distinct RBs from the feasible RB pool catered to them.

\begin{figure}
\centering
\raisebox{-.5\height}{%
  \includegraphics[width=3.5in]{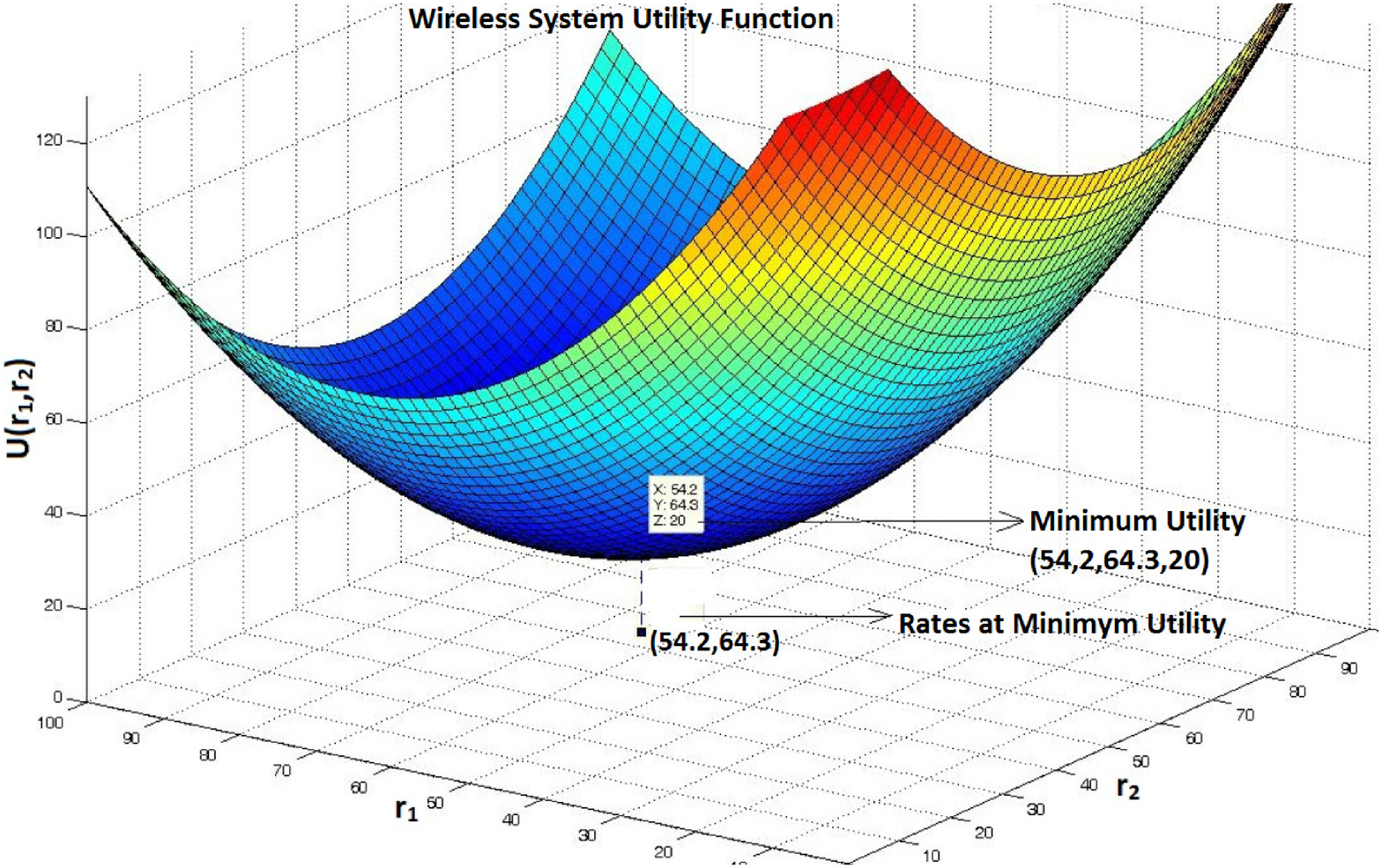}%
}\qquad
\raisebox{-.5\height}{%
  \includegraphics[height=4cm,width=3.5in]{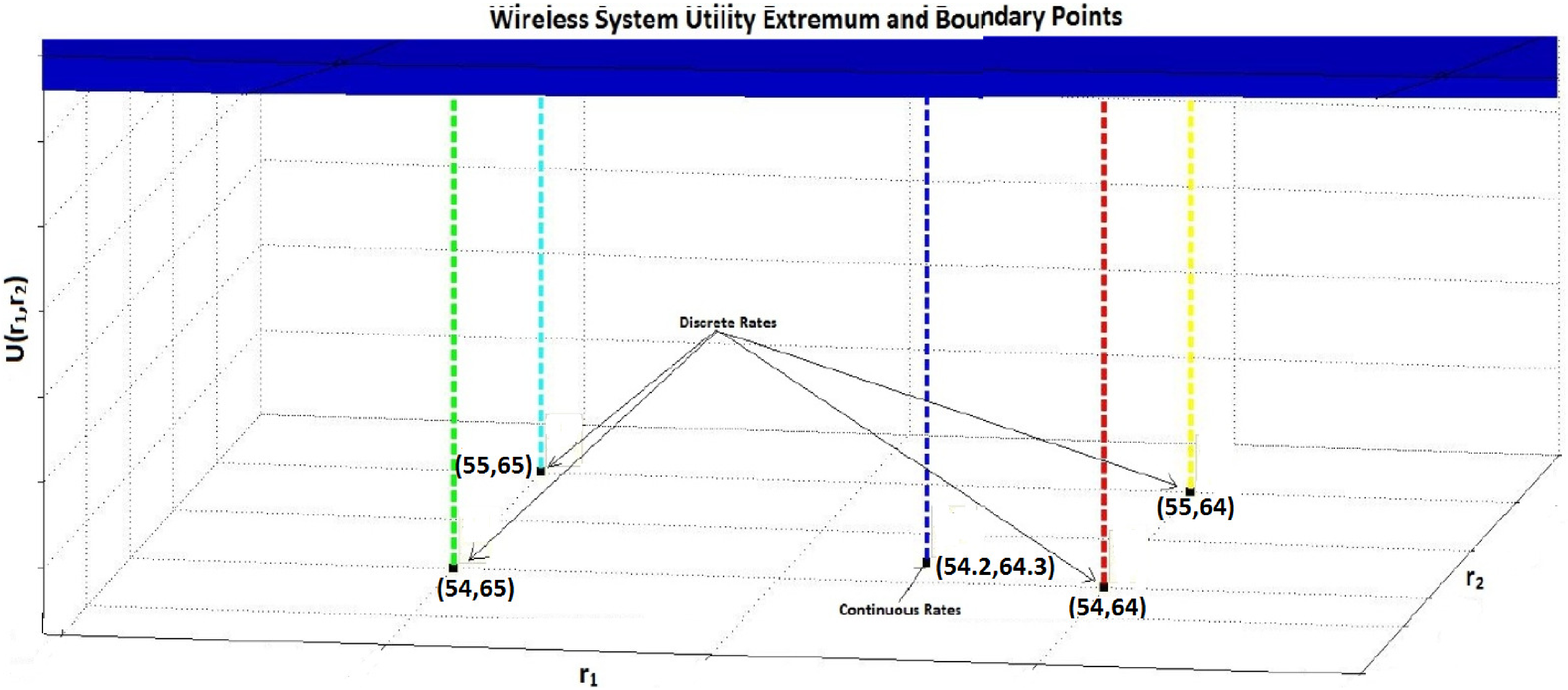}%
}
\caption{RB Allocation Notion: Atop, a two-UE system with the utility $U(r_1,r_2)=U(r_1)U(r_2)$ and rates $r_1$ and $r_2$ has a minimum $U(r_1,r_2) =20$ at $(r_1,r_2) = (64.3,54.2)$, the optimal continuous rates displayed as the blue point at the bottom. The red, green, yellow, cyan points $(r_1,r_2)$ = $(54,64)$, $(54,65)$, $(55,64)$, $(55,65)$ are the discrete rates the closest to the blue optimal point.}
\label{fig:UtilityExample2D}
\end{figure}

\section{Simulation Results}\label{sec:sim}
\begin{figure}[!htb]
\begin{center}
\includegraphics[width=3.5in]{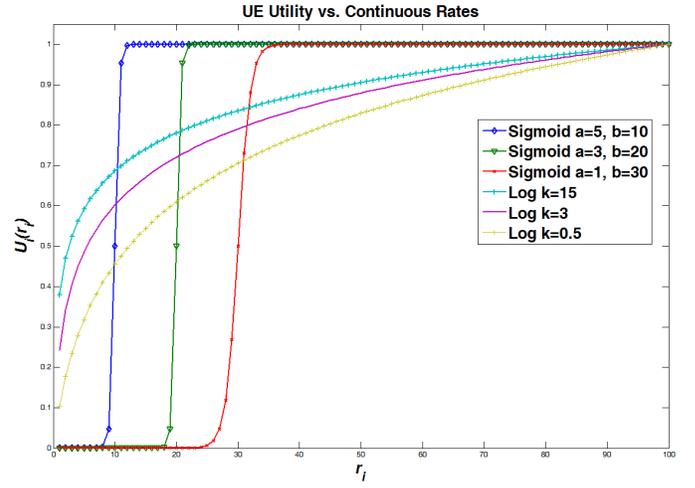}
\end{center}
\caption{Utility Functions: The abscissa and ordinate represent continuous rates and UE utility respectively. Cyan, purple, and yellow are delay-tolerant FTP applications with logarithmic utility with $k = 15, 3, 0.5$ respectively. Red, green, and blue are real-time HDV, standard video, VoIP with sigmoidal utility $(a,b) = (1,30),(3,20), (5,10)$ respectively.} \label{fig:ContUtility}
\end{figure}

Algorithms (\ref{alg:Our_UE}) and (\ref{alg:Our_eNodeB}) were applied in MATLAB to the scenario in Fig. \ref{fig:SysMod} where UEs 1, 2, and 3 run real-time applications with the sigmoidal utility parameters $(a,b) = (5,10), (3,20), (1,30)$ for a Voice-over-IP (VoIP), standard streaming video, and high definition TV (HDV) application and UEs 4, 5, and 6 run delay-tolerant FTP services with $k = \{15, 3, 0.5\}$, depicted in Fig. \ref{fig:ContUtility}. The red HDV has the highest inflection point $30$, whereas the green standard video and blue VoIP have respectively lower inflections points $20$ and $10$ implying they need less QoS resources, and an inflection-exceeding allocation saturates $100\%$ utility. The cyan, purple, and yellow logarithmic utilities are FTP applications in an ascending delay-tolerance order, e.g. at identical rates, the yellow utility is under the other logarithmic ones manifesting its less delay-tolerance.

\begin{figure}
\centering
\raisebox{-.5\height}{%
  \includegraphics[height=4cm,width=3.5in]{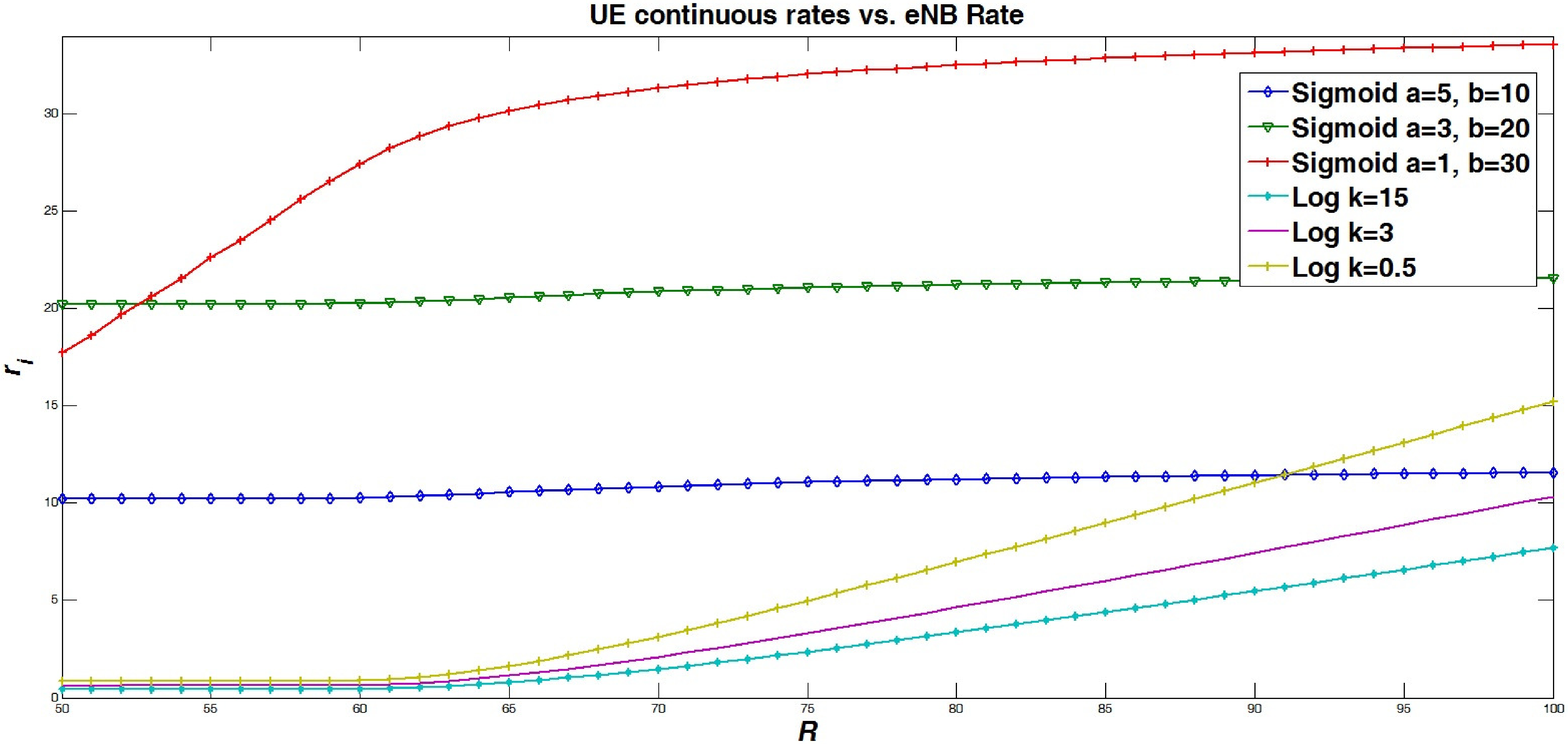}%
}\qquad
\caption{eNB Rate $R$ vs. Continuous Rate for the Relaxed Problem: The eNB $R$, the higher UE continuous rates. Red, green, and blue (Cyan, purple, and yellow) curves are rates for the UEs 1, 2, and 3 (4, 5, and 6) running VoIP, standard streaming video, and HDV (FTP) applications. Real-time applications are initially allotted more resources due their higher QoS needs.}\label{fig:ConRatR}
\raisebox{-.5\height}{%
  \includegraphics[width=3.5in]{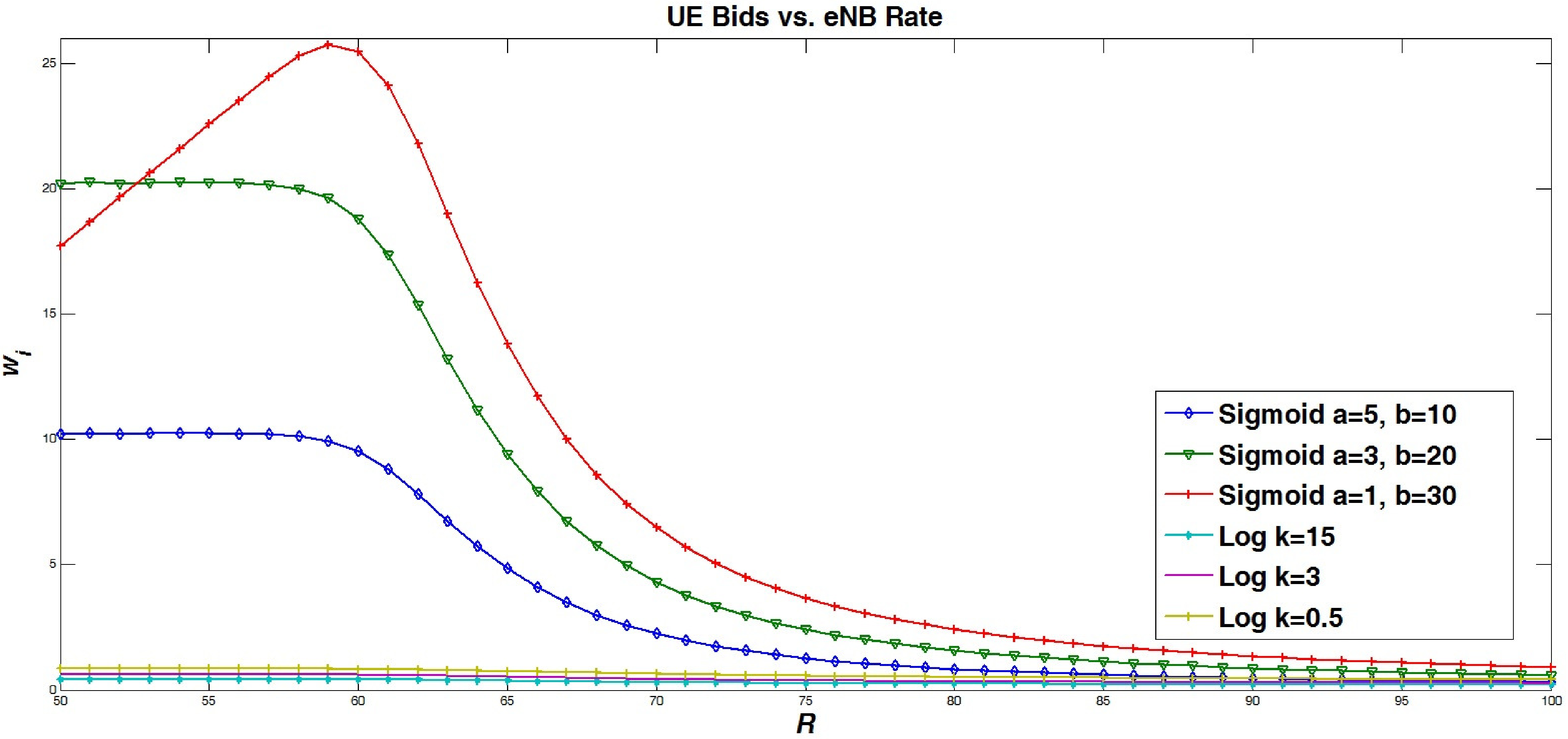}%
}
\caption{eNB Rate $R$ vs. Bids for the Relaxed Problem: The eNB $R$, the higher UE continuous rates, the lower bids. Real-time applications initially (scarce resources) bid higher to earn resources to meet their minimum QoS needs.} \label{fig:UEBidsR}
\label{fig:UtilityExample2D}
\end{figure}

We set the algorithm iterations to $40$ and range the eNB rate $R$ from $50$ to $100$ in unit steps. In Fig. \ref{fig:ConRatR}, we observe that our approach does not allocate UEs any zero rates signifying that no user is dropped and is a direct result of employing a proportional fairness formalization. Moreover, real-time applications bandwidth assignment continues until reaching their inflection points, and if $R$ falls short of the inflection points sum, real-time applications get less resources than the inflection points which gravely impacts the QoS; however, an $R$ exceeding the inflection point sum $\sum b_i$ entails more allocations to the delay-tolerant applications as well; this situation occurs when $R$ surpasses $60$ in our simulation. In addition, we look UE bids on resources whose availability alters at the eNB in Fig \ref{fig:UEBidsR}. The larger bids are tantamount to higher allocated rates, and more eNB resources ensue less bandwidth scarcity, thereby it slashes down UE bids severely.

\begin{figure}[!htb]
\begin{center}
\includegraphics[width=3.5in]{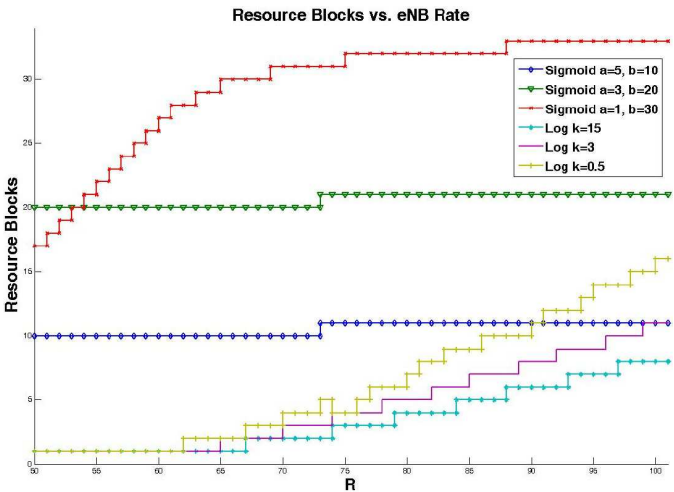}
\end{center}
\caption{Resource Block Allocation versus eNB Rate $R$ for the Discrete Optimization Problem: Red, green, and blue (cyan, purple, and yellow) curves are RBs for the UEs 1, 2, and 3 (4, 5, and 6) running VoIP, streaming video, and HDV (FTP) applications respectively. Other RBs are also feasible.} \label{fig:RBAllR}
\end{figure}

Now that the relaxed problem is solved, the RB allocation is performed as in Fig. \ref{fig:RBAllR}. The structure of many cellular networks' resource allocation such as the LTE relies on RBs, so the RB assignment make any resource allocation strategy pragmatic. As we can observe, UEs 1, 2, and 3 running real-time applications get hold of more RBs at the commencement when resources are limited. In contrast, enlarging $R$ improves the resource availability and affords the eNB to assign further RBs to the UEs with delay-tolerant applications. To make the process more tangible, we look into exemplar continuous rate and RB allocations for the system in Fig. \ref{fig:SysMod} with arbitrarily selected eNB bandwidth $R$ enumerated at the first column of table \ref{table:Rates}, whose second column is a $M = 6$ dimensional vector whose $i^{th}$ element is the $i^{th}$ UE continuous optimal rate obtained from the relaxed optimization problem and last column is the corresponding RB allocations.

Apparently, an optimal continuous rate assignment might be mapped to one or more RB candidates proffering the eNB a flexibility to allocate freely from an RB feasibility pool. As a case in point, for $R = 100$, the relaxed problem allocates the UEs the optimal rates [$11.57,21.57,33.58,7.72,10.36,15.21$] equivalent to $8$ RB allocations feasibilities (Thirst column of table \ref{table:Rates}) whereas $R = 50$ leads to the optimal rate [$10.46,20.46,24.21,9.13,9.13,9.13$] and one RB allocation alone. This comes handy where several eNBs work in a cellular environment with intercell and intracell interference detrimental to the user QoE. Nonetheless, such analysis is not investigated in the current article.

\begin{table}[h]
\caption{An Example of Continuous Optimal Rates vs. the RB Assignments: eNB Rate $R = 50$ ($100$) yields in one (eight) RB candidate(s).}
\centering
\begin{tabular}{c c c}
\hline\hline
R & $r_i$ & RB \\[0.5ex]
\hline
100 & $11.57,21.57,33.58,7.72,10.36,15.21$ & $11,21,33,8,11,16$\\
100 & $11.57,21.57,33.58,7.72,10.36,15.21$ & $11,21,33,8,11,15$\\
100 & $11.57,21.57,33.58,7.72,10.36,15.21$ & $11,21,33,8,10,16$\\
100 & $11.57,21.57,33.58,7.72,10.36,15.21$ & $11,21,33,8,10,15$\\
100 & $11.57,21.57,33.58,7.72,10.36,15.21$ & $11,21,33,7,11,16$\\
100 & $11.57,21.57,33.58,7.72,10.36,15.21$ & $11,21,33,7,11,16$\\
100 & $11.57,21.57,33.58,7.72,10.36,15.21$ & $11,21,33,7,10,16$\\
100 & $11.57,21.57,33.58,7.72,10.36,15.21$ & $11,21,33,7,10,15$\\
50  & $10.46,20.46,24.21,9.13,09.13,9.13$   & $10,20,17,01,01,01$\\[1ex]
\hline
\end{tabular}
\label{table:Rates}
\end{table}

\begin{figure}[!htb]
\begin{center}
\includegraphics[width=3.5in]{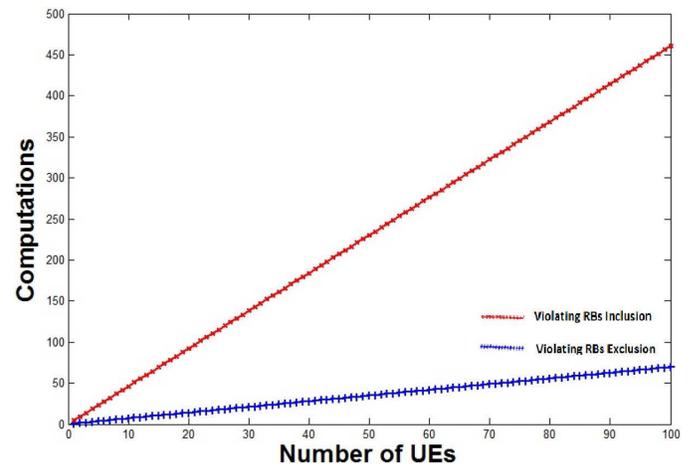}
\end{center}
\caption{Computational Complexity: Red curve shows the semi logarithmic computational complexity as a function of the number of UEs once we do not exclude violating RB possibilities, while the blue curve is the semi logarithmic computational complexity once we have applied the technique to exclude the RBs infringing the constraint for the RB allocation problem.} \label{fig:ComCompl}
\end{figure}

Furthermore, it is worthwhile to study the computational complexity and execution runtime associated with the proposed scheme. Since for every continuous optimal rate $r_i$ allocated to the $i^{th}$ UE with the utility $U_i$, we have at least one and at most two possible discrete boundary points, the computational complexity will be at most $O(2^{M})$ for an $M$ UE network. That is to say not all of the $2^{M}$ lower and upper bounds of the continuous optimal rates satisfy the relaxed problem's constraint requiring an RB augmentation not larger than the eNB bandwidth. In conclusion, some potential RB candidates are omitted due to infringing the eNB rate criteria in the RB allocation problem. Exclusion of the violating RBs is a significant remedial measure for the computational complexity. As an illustration, assuming $n$ RB possibilities per UE, an $M$-UE system's computational complexity reduces from $O(n^{M})$ to $O(2^{M})$ at the outside. That is to say an eNB with $10$ RB candidates covering $100$ UEs 
observes a plummet from $10^{200}$ to solely $2^{100}$ computations, depicted in Fig. \ref{fig:ComCompl} as a semi logarithmic diagram in view of the exponential increase in the number of computations vis-a-vis the UE quantity. On the other hand, the runtime of the RB allocation for our simulation is relatively short so that when the eNB rate changes from $50$ to $100$ bandwidth units, the runtime was 16 seconds. This is reasonable as it may take a much larger time for the eNB bandwidth (resources available to the base station by the network provider) to change $51$ units abruptly. Besides, the RB allocation algorithm runtime for fixed eNB rates was $0.24$ - $0.30$ seconds in MATLAB.

\section{Conclusion}\label{sec:conclude}
In this paper, we studied a resource block allocation optimization problem cast under a utility proportional fairness framework to assign resources to user equipments in a cellular infrastructure based on radio resource blocks with an attention to quality of service requirements for real-time/delay-tolerant applications represented by sigmoidal/logarithmic utility functions. Initially, we formalized the resource block allocation as an integer optimization difficult to solve. Then, we leveraged a Lagrangian relaxation to transform the discrete optimization problem into a continuous rate allocation convex optimization problem, solved via Lagrange multipliers of the dual problem to engender continuous optimal rates.

Then, a boundary mapping of the continuous rates to discrete ones below and above the optimal rates was adopted. This technique yielded in a series of candidate discrete rates representing potential resource blocks for the user equipments. We fine-grained the selection by choosing resource blocks satisfying the relaxed continuous problem constraints and further evaluated the constraint-met resource blocks to locate those which maximized the cellular system's proportional fairness utility function. The resultant resource blocks constitute a feasible discrete rate set affording the base station the luxury of boasting several resource block assignments possibilities per individual optimal continuous rate assignment.

Besides, we investigated the sensitivity of our algorithm to the available resources mutations, obtained user bids, continuous rates, and resource blocks, and realized the less scarce resources, the lower the user bids, and the more resource blocks allocations. In addition, we pointed out that the proposed scheme rendered user equipments with real-time applications higher priorities, thereby it delivered the application quality of service. Furthermore, we realized that for base station bandwidths surpassing the utility inflection points sum, resources can be assigned delay-tolerant applications generously. Ultimately, we investigated the complexity of the algorithm and realized that the problem is computationally efficient and has a reasonably short runtime. 

\bibliographystyle{ieeetr}
\bibliography{Pub}
\end{document}